\documentclass[preprint,showpacs,preprintnumbers,amsmath,amssymb]{revtex4}

\usepackage{graphicx}
\usepackage{dcolumn}
\usepackage{bm}

\begin{document}

\preprint{}

\title{A simple algorithm to test for linking to Wilson loops in percolation}

\author{Robert M. Ziff}
 
\affiliation{%
Michigan
Center for Theoretical Physics, and\\ 
Department of Chemical Engineering\\ University of Michigan,\\ Ann Arbor, MI 48109-2136 USA\\}%

\date{\today}

\begin{abstract}
A simple burning or epidemic type of algorithm is developed in order
to test whether any loops in percolation clusters link a fixed reference
loop, a problem considered recently by Gliozzi, Lottini, Panero,
and Rago in the context of gauge theory.   We test our algorithm at
criticality in both 2d, where
the behavior agrees with a theoretical prediction, and in 3d.
\end{abstract}

\maketitle

\section{\label{sec:level1}Introduction}
Recently, Gliozzi et al.\ \cite{Gliozzi05,Gliozzi04} have studied percolation 
in the context of gauge theory.  They considered the question of whether
closed paths in three-dimensional percolation clusters are linked topologically to 
given closed loops, the so-called Wilson loops.   Studying this
problem for rectangular planar loops,
and in comparison to percolation in three-dimensional slabs which they relate to 
the problem of deconfinement,  the authors find a new universal amplitude ratio.
This work provides an example where 
the percolation model possesses connections to fundamental problems in
theoretical particle physics.

Gliozzi et al.'s numerical
results for rectangular loops of dimensions $R \times T$
confirmed the expected behavior  for $p \ne p_c$ \cite{Ambjorn84}
\begin{equation}
\langle W(R,T)\rangle=C\,e^{-P(R+T)-\sigma RT}
\, R^{1/4} \sqrt{\frac{\eta(i)}{\eta(iT/R)}}~,
\label{weq}
\end{equation}
where $\langle W(R,T) \rangle$ is the average probability that there is no path in 
any cluster linked to the Wilson loop, $C, P$, and $\sigma$ are constants that 
depend upon the percolation probability $p$, and $\eta$ 
is the Dedekind function
$\eta(\tau)=q^{1/24}\prod_{n=1}^\infty(1-q^n)$ with $q=e^{2i\pi\tau}$.
When $p < p_c$, one expects $\sigma = 0$ because the
linking probability should depend only upon the perimeter of the loop, while for
$p > p_c$, the linking probability is expected to decay
exponentially with the area of the loop ($\sigma > 0$).
Taking $p$ somewhat above $p_c$, Gliozzi et al.\ found
that the dependence of $\sigma$ upon $p$ behaves as 
\begin{equation}
\sigma=S\,(p-p_c)^{2\nu}~~,
\label{sscaling}
\end{equation}
similar to the behavior of a surface tension,
where $\nu \approx 0.8765$ is the correlation-length exponent 
of 3-d percolation \cite{Ballesteros99}, and $S$ is a constant.  
They also determined the percolation threshold $p_\ell$ for slabs of
thickness $\ell$ in the range 3 -- 8. One expects
 $\ell \sim \xi(p_\ell) \sim T_c^{-1} (p_\ell - p_c)^{-\nu}$ where $T_c$ is a constant, 
and indeed they find $1/(\ell \sqrt{\sigma(p_\ell)}) \sim T_c / \sqrt{S} $ is a universal amplitude
ratio with a value of about 1.50.

In this note, we discuss two points related to the work of Gliozzi et al.:  (1) We describe an epidemic
or burning \cite{StaufferAharony}  type of algorithm that may be simpler than the algorithm described
by Gliozzi et al., and (2) we apply it to study the linking probability
exactly at $p_c$ (a point that Gliozzi et al.\ did not consider) for 2-d and 3-d systems.
Note that Eq.\ (\ref{weq}) and its 2-d analog are not necessarily expected
to be valid at $p_c$. 

\section{Algorithm}

Gliozzi et al.\ describe an algorithm that involves successive removal of dangling ends
and reduction to an auxiliary graph that represents the connections between clusters
on either side of the flat region enclosed by the loop.
This graph is used to determine whether a cluster is linked to the loop.

Here we describe a cluster burning type of algorithm that accomplishes the same
test.  As in Ref.\ \cite{Gliozzi05}, we consider the loop $\gamma$ to be on the dual lattice, so
effectively the problem is to find if there are clusters that simultaneously pass 
through the plane $\Sigma$ of vertical bonds enclosed by $\gamma$ and
through bonds in the same plane outside of $\gamma$.  We are
thinking of a simple cubic lattice to be specific.

To begin the process, all of the sites are set to the ``unvisited" 
state and bonds to the ``undetermined" state.  Then we pick
one of the (unvisited) sites ${\cal S}$ directly above $\Sigma$, and label that site
as ``visited" with an arbitrary index $n$.   We check the
six bonds that emanate from ${\cal S}$; the undetermined bonds 
are made ``occupied" with probability $p$ and ``vacant" 
otherwise.  For the bonds that are occupied, we check the adjacent site;
if that site is unvisited, we label it as visited (with a value of the label
described below) and put
its coordinates on a queue for future checking. After finishing checking all bonds
connected to the site being studied, we consider the next site on the
queue, continuing this process until
the queue is empty.  This is the normal burning or epidemic type of algorithm to
identify a cluster connected to a site in bond percolation; here we also decide
whether a bond is occupied or not as we go along.  We repeat this process for
all remaining unvisited sites in the plane above $\Sigma$.

What we now do differently for the loop-linking problem is that we assign
an index $n$ to each visited site in a cluster.  When we
transverse one
of the occupied bonds that intersects $\Sigma$, we increment
$n$ by one when going downward or decrement it by one when going
upwards.
In this way, every site of the cluster will be
labelled by $n$, $n\pm 1$ (if a path of the growing cluster
goes once through $\Sigma$), $n\pm 2$
(if a path of the growing cluster winds twice through $\Sigma$), etc.

Now, if during the growing process a new bond is found to connect two 
visited sites with
different labels ${\cal L}$, then the cluster must have wrapped around
$\gamma$ and is therefore linked to it.

An occupied bond of course will not connect to sites of two different clusters,
by definition, so therefore one does not have to worry about interference
between clusters in this algorithm.

Here we have not dealt with the system boundaries.  If open boundaries
are used, care must be taken so that the boundary bonds are not
mistaken for wrap-arounds.  To eliminate this problem and to lessen the
effects of the boundary, we considered periodic boundary conditions
in all directions.  This 
makes the problem slightly different, because periodic
wraparounds through the 
Wilson loop will also contribute
to linking events,
but if the lattice dimension $L \gg R$, this difference should not be too significant.  Indeed, a moment's
reflection shows that it is very unlikely that there will be wraparound without linking,
since a cluster that wraps around the lattice it is usually a
ubiquitous one and most likely will link the Wilson loop also.

The idea of adding a label to sites in percolation to test for
a crossing criterion has been used previously in relation to wrapping
a periodic system in a given direction \cite{Machta,NewmanZiff01}.

\begin{figure}
\includegraphics[width=4in]{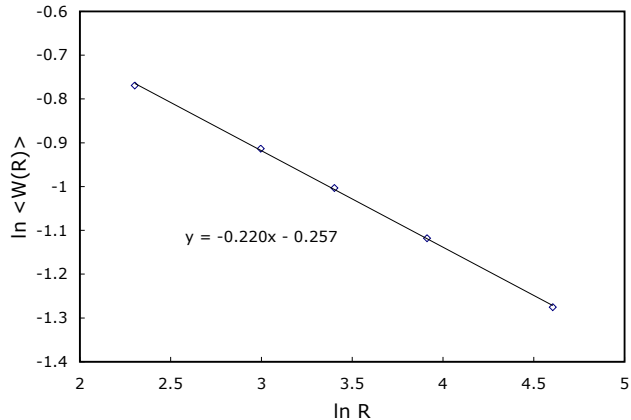}
\caption{\label{2dfig}Logarithm of $\langle W(R) \rangle$ (=  the probability that no cluster encircles
just one of the two points), vs.\ the logarithm of the points' separation $R$.}
\end{figure}

\section{Linking in two dimensions at criticality}

For
the 2-d system, the question that is studied is whether there exists a closed
path in a cluster that encircles one (but not both) of two points
on the dual lattice, separated
by a distance $R$.  In this case
we can make a simple theoretical prediction for $\langle W(R) \rangle$, 
since the condition of a path not encircling either of the two points
individually  is equivalent to the 
existence of a continuous path between the two points on the dual lattice --- that is,
a cluster that connects the two points.  The density drop-off from
any point on a given cluster goes as
$r^{D-d}$ where $D$ is the fractal dimension and
$d$ is the spatial (Euclidean) dimension.  To find the probability that two
given points are connected, the above factor must be multiplied by the
probability the size of a cluster connected to one of the points is at least
large enough to reach the other point.  At criticality, the probability that the number of
sites connected to a point
is equal to or greater than $s$ is given by  $P_{\ge s} \sim s^{2-\tau}$ where $\tau$
is the size distribution exponent, and this implies that the probability that the radius is
greater than or equal to $r$ is given by $P_{\ge r} \sim r^{D(2 - \tau)}$, since $s \sim r^D$.
Then, by the hyperscaling relation $d/D= \tau - 1$, we have  $P_{\ge r} \sim r^{D-d}$. 
Thus, the net probability that two points separated by $r$ are connected by a cluster
at $p_c$ is given by $P(r) \sim r^{2(D-d)}$, implying that
\begin{equation}
\langle W(R) \rangle \sim R^{-2(d-D)} = \exp( -2(d-D) \ln R )
\label{W2d}
\end{equation}
In $d = 2$, $D = 91/48$ and $2(d-D) \approx 0.208$.

We carried out simulations for this system using 
the algorithm described above.  We considered bond
percolation on the square lattice at $p = p_c = 1/2$, on a system with a square boundary
of dimensions $1024 \times 1024$, and considered separations
of the two points ranging between 10 and 100.  
Fig.\ 1 shows the results for a plot of $\ln W$ vs.\ $\ln R$, 
for a relatively small number of runs (100000 each).  The slope is
about $-0.22$, consistent with the theoretical prediction above.
To make this work more precise, one would have to consider
different size systems to study the finite size corrections, and
perhaps also consider systems with open boundary conditions
for comparison.

\begin{figure}
\includegraphics[width=4in]{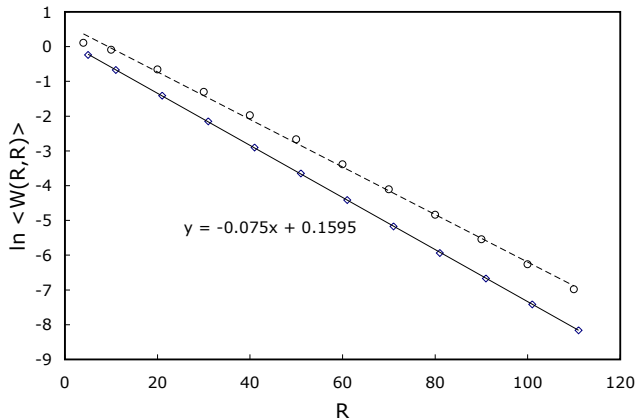}
\caption{\label{3dfig} 3-d data. Upper ($\circ$): $\ln (R^{-1/4} \langle W(R,R)\rangle$) vs.\ $R$.
Lower ($\diamond$): $\ln  \langle W(R,R) \rangle$ vs.\ $R$, which shows a good
fit to a straight line for $5 \le R \le 111$.  The equation of the linear fit is given.}
\end{figure}

\section{Linking in three dimensions at criticality}

For the 3-d problem, we consider bond percolation on the simple
cubic lattice, and take $p = 0.2488126$, which is an estimate for 
$p_c$ believed to
be within about $5 \cdot 10^{-7}$ of the actual value \cite{LorenzZiff98}.  We consider a lattice
of size $128 \times 128 \times 128$, and square Wilson loops containing
$R \times R$ vertical bonds, with $R = 5, 11, 21, \cdots, 111$.  Between 300,000
(smaller $R$) and 13,000,000 samples (larger $R$) were generated for the different 
values of $R$.

In Fig.\ 2, the lower curve represents
$\ln \langle W(R,R) \rangle$ as a function of $R$.  The data 
shows quite linear behavior up to $R = 111$.  Evidently, the perimeter term
proportional to $P$ in Eq.\ (\ref{weq}) dominates; as expected,
there is no  term proportional to the area.

We do not see evidence of the $R^{1/4}$ term in Eq.\ (\ref{weq}).
The data marked by circles in Fig.\ 2 represents $\ln (R^{-1/4} \langle W(R,R) \rangle)$
vs.\ $R$, and the fit to a straight line is much worse than for the case
without the factor of $R^{-1/4}$.  This factor would show up as a logarithmic
term in the plot of Fig.\ 2.

\begin{figure}
\includegraphics[width=4in]{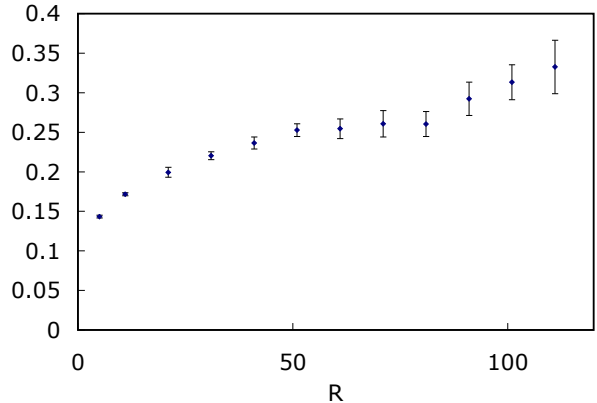}
\caption{\label{3dfig2} $\ln W+ 0.0765 R$ vs. $R$ for the 3-d data,
showing deviations from simple exponential behavior.
Error bars show two standard deviations of statistical error.}
\end{figure}

To check further for logarithmic terms, we plot in Fig.\ 3 the quantity $\ln W+ 0.0765 R$,
where the constant 0.0765 was adjusted to get the best horizontal region in the
center, along with general monotonic behavior.  We see two corrections to the
straight line: for small $R$, there is a small decrease, which could
be fit to a very small logarithmic term, $\approx -0.03 \ln R$, much smaller than the
$-(1/4) \ln R$ term that would appear for $p > p_c$ according to Eq.\ (\ref{weq}).
The coefficient is so small that the existence of a logarithmic term seems unlikely.

For large $R$, the deviations
from linearity are also small, which is surprising given that we went up to $R=111$
in a system of size $L = 128$.  When $R$ approaches $L$ it should be more
difficult to create a linking clusters, because there is a smaller region external to
the loop (also taking into account the periodic b.c.), but this may be balanced by
the increase in vertical wraparounds (which our algorithm takes to be linkages) through the
periodic boundary conditions.  Note that at $R = 111$, $W = 0.00028617$ --- that is,
only 3434 of the 12,000,000 samples did not have a linkage to the Wilson loop.

Thus, ignoring the possible small logarithmic term, the
 data for 3-d (in the central range) yields $ \ln W = - 0.0765 R + 0.25$, implying
\begin{equation}
\langle W(R,R)\rangle=1.28\,e^{-0.0382 (2R)}
\label{W3d}
\end{equation}
or $P = 0.0382$ for bond percolation on the simple cubic lattice at  criticality.
Note the linear fit for $W$
above is somewhat different than that given in Fig.\ 2, which is
just a simple linear fit through all the data points.

\section{Conclusions}

We see that a simple burning type of algorithm can be constructed
to find the loop linking probability studied by Gliozzi et al.  We have
checked it in two dimensions at the critical threshold, where the
linking probability is known exactly by virtue of its being dual
to the two point probability.  Of course, in 2d one can
easily simulate the dual problem of connecting the two points.
However, in 3d, where a
dual-lattice procedure would be much more complicated,
a direct determination is preferable and
the algorithm presented here is efficient and simple.
For 3d, we find a simple exponential relation between
$\langle W(R,R) \rangle$ and $R$ reflecting a perimeter
effect; there is no evidence of a logarithmic correction implied
by the $R^{1/4}$ term in Eq.\ (\ref{weq}) (which is not necessarily
expected to be valid at $p_c$) or as suggested by the behavior in 2d.

\section{Acknowledgements}
The author acknowledges support of the
National Science Foundation under grant number DMS-0244419.

\end{document}